\documentclass[conference]{IEEEtran}
\IEEEoverridecommandlockouts
\usepackage{cite}
\usepackage{hyperref}

\usepackage{amsmath,amssymb,amsfonts}
\usepackage{algorithmic}

\ifCLASSINFOpdf
    \usepackage[pdftex]{graphicx}
    \DeclareGraphicsExtensions{.pdf,.jpeg,.png}
\else
  \usepackage[dvips]{graphicx}
\fi

\ifCLASSOPTIONcompsoc
  \usepackage[caption=false,font=normalsize,labelfont=sf,textfont=sf]{subfig}
\else
  \usepackage[caption=false,font=footnotesize]{subfig}
\fi

\usepackage{textcomp}
\usepackage{xcolor}

    \def\Complex{{\rm\rule[.23ex]{.03em}{1.1ex}\kern-.3em{C}}}

    \newcommand{\be}{\begin{equation}} \newcommand{\ee}{\end{equation}}
    \newcommand{\bea}{\begin{eqnarray}} \newcommand{\eea}{\end{eqnarray}}
    \newcommand{\benum}{\begin{enumerate}} \newcommand{\eenum}{\end{enumerate}}

    \newcommand{\qd}{{\bf d}}
    
    \newcommand{\qf}{{\bf f}}

    \newcommand{\qr}{{\bf r}}

    \newcommand{\qB}{{\bf B}}

    \newcommand{\qP}{{\bf P}}

    \newcommand{\qT}{{\bf T}}

    \newcommand{\bbR}{{\mathbb R}}
    \newcommand{\bbC}{{\mathbb C}}

    \usepackage{xcolor}

    \newcommand{\rl}[1]{\color{red}#1}

\def\BibTeX{{\rm B\kern-.05em{\sc i\kern-.025em b}\kern-.08em
    T\kern-.1667em\lower.7ex\hbox{E}\kern-.125emX}}

\begin{document}
\title{Cross-Band Channel Impulse Response Prediction: Leveraging 3.5 GHz Channels for Upper Mid-Band
}

\author{
\IEEEauthorblockN{Fan-Hao~Lin\IEEEauthorrefmark{1},
 Chi-Jui~Sung\IEEEauthorrefmark{1},
 Chu-Hsiang~Huang\IEEEauthorrefmark{2},
 Hui~Chen\IEEEauthorrefmark{3},
 Chao-Kai~Wen\IEEEauthorrefmark{1},
 Henk Wymeersch\IEEEauthorrefmark{3} }
       
\IEEEauthorblockA{\IEEEauthorrefmark{1}Institute of Communications Engineering, National Sun Yat-sen University, Kaohsiung 80424, Taiwan} 

\IEEEauthorblockA{\IEEEauthorrefmark{2}Department of Electrical Engineering, National Taiwan University, Taipei 10617, Taiwan}

\IEEEauthorblockA{\IEEEauthorrefmark{3}Department of Electrical Engineering, Chalmers University of Technology, 41296 Gothenburg, Sweden}
}
\maketitle

\begin{abstract}
Accurate cross-band channel prediction is essential for 6G networks, particularly in the upper mid-band (FR3, 7--24 GHz), where penetration loss and blockage are severe. Although ray tracing (RT) provides high-fidelity modeling, it remains computationally intensive, and high-frequency data acquisition is costly. To address these challenges, we propose \textbf{CIR-UNext}, a deep learning framework designed to predict 7 GHz channel impulse responses (CIRs) by leveraging abundant 3.5 GHz CIRs. The framework integrates an RT-based dataset pipeline with attention U-Net (AU-Net) variants for gain and phase prediction. The proposed AU-Net-Aux model achieves a median gain error of 0.58 dB and a phase prediction error of 0.27 rad on unseen complex environments. Furthermore, we extend CIR-UNext into a foundation model, \textbf{Channel2ComMap}, for throughput prediction in MIMO-OFDM systems, demonstrating superior performance compared with existing approaches. Overall, CIR-UNext provides an efficient and scalable solution for cross-band prediction, enabling applications such as localization, beam management, digital twins, and intelligent resource allocation in 6G networks.
\end{abstract}


\section{Introduction} 
Future 6G networks aim to integrate the entire spectrum to deliver high-throughput and low-latency services envisioned in the IMT-2030 framework \cite{itu2023imt2030framework}. This includes the newly considered upper mid-band (FR3, ~7--24 GHz) \cite{5GAmericasFR3}. Reliable communication in FR3 demands accurate channel estimation and prediction, yet the scarcity of measurement datasets at these frequencies poses major challenges. Consequently, efficient channel prediction has become essential for enabling advanced functions such as digital twins, fingerprint-based localization, and intelligent beam management.

Conventional approaches rely on ray tracing (RT) \cite{aoudia2025sionna}, which deterministically models reflections, diffractions, and scatterings. Although RT provides physically accurate channel representations at the target frequency, it remains computationally prohibitive, particularly for large-scale or multi-band scenarios \cite{li2025coverageprediction}. Rich channel data are already available in the lower mid-band (e.g., 3.5~GHz), which exhibits strong geometric correlations with upper mid-band channels. This naturally raises a key question: \textit{can low-band channel impulse responses (CIRs) be leveraged to infer high-band CIRs without repeating costly measurements or extensive RT simulations?}  

Recent studies have explored AI-driven cross-frequency channel and throughput prediction. Generative models have been employed to infer statistical channel features across frequency bands \cite{Hu-24SPAWC}, while RadioTwin \cite{an2025radiotwin} learns material-dependent electromagnetic properties to achieve frequency-consistent modeling. Geo2ComMap \cite{lin2025geo2commap} performs cross-band mapping by leveraging sparse throughput indicators and geographic information, and \cite{Chen25RatePrediction} applies a Transformer network for multi-band and temporal rate prediction. In addition, physics-informed neural networks (PINNs) have been introduced for channel modeling, integrating electromagnetic principles to enhance generalizability and interpretability \cite{Zhu2024PINN}. However, these studies do not explicitly learn per-path cross-band mappings that jointly predict both path gain and phase, and thus remain limited to statistical- or throughput-level inference.

To address this gap, this paper proposes \textbf{CIR-UNext}, a deep learning framework for cross-band channel prediction that transfers multipath information from 3.5 GHz to 7 GHz. The main contributions of this work are summarized as follows:

\begin{itemize}
    \item We propose \textbf{CIR-UNext}, a deep learning framework that jointly predicts multipath gain and phase across frequency bands, enabling accurate path-level channel reconstruction from 3.5 GHz to 7 GHz.
    
    \item The framework employs specialized Attention U-Net (AU-Net) architectures for multipath gain and phase prediction. The AU-Net integrates attention gates (AGs) \cite{oktay2018attention} into the U-Net skip connections to enhance salient regions while suppressing irrelevant features. 
    
    \item Extensive experiments identify key auxiliary inputs, such as antenna patterns, angles of departure (AoD), and delay information, that substantially improve model robustness, particularly under complex non-line-of-sight (NLoS) conditions.

    \item We further introduce \textbf{Channel2ComMap}, which extends CIR-UNext as a foundation model for throughput prediction in 5G MIMO-OFDM systems. The proposed model significantly outperforms the state-of-the-art Geo2ComMap \cite{lin2025geo2commap}, which relies on sparse throughput samples and geographic information. 
 
\end{itemize}

\begin{figure*}[!t]
    \centering
    \includegraphics[width=0.85\textwidth]{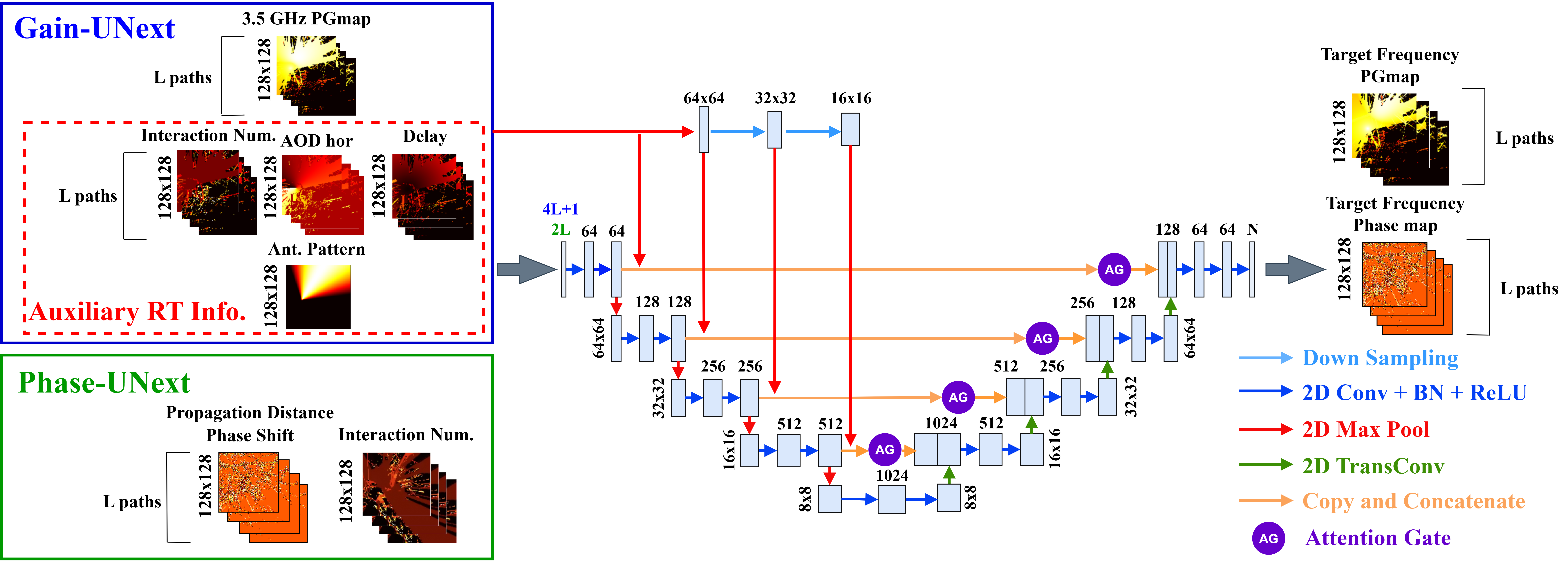}
    \caption{Architecture of \textbf{CIR-UNext}: \textbf{Gain-UNext} is an AU-Net-Aux with a $(4\times L + 1)$-channel input for efficient and accurate per-path gain prediction, where $L$ is the number of resolved multipath components. \textbf{Phase-UNext} shares the simliar AU-Net backbone (without the {\color{red} auxiliary part}), with a $(2\times L)$-channel input, but is trained separately for per-path phase prediction. In our experiments, we consider $L=20$.}
    \label{fig:Gain-UNext}
\end{figure*}

\section{Framework Design} \label{System Design}
This section presents the methodology of \textbf{CIR-UNext}, which consists of two main stages. The first stage constructs ray-tracing datasets at both 3.5~GHz and 7~GHz using geographical building maps. The second stage employs a supervised deep learning framework to learn an efficient cross-band mapping from the 3.5~GHz channel to its 7~GHz counterpart. The overall design is motivated by the physical invariance of multipath geometry across frequencies, as detailed below.  

\subsection{Physical Motivation for Cross-Band Inference}

The feasibility of cross-band inference can be understood through the 3GPP channel model. According to 3GPP TR~38.901 \cite{3gpp38901} (see also \cite{aoudia2025sionna}), the CIR from the $s$-th transmit (Tx) antenna to the $u$-th receive (Rx) antenna at frequency $f$ is expressed as
\begin{equation} \label{eq:h_su}
    h_{s,u}(\tau,f) = \sum_{l=1}^{L} h_{s,u,l}(f)\,\delta(\tau-\tau_l),
\end{equation}
where $L$ is the number of multipath and $\tau_l$ is the delay of the $l$-th path. Each per-path term can be modeled as
\begin{multline} \label{eq:h_sul}
    h_{s,u,l}(f) \!=\!
    \qf_{\text{rx}}^{T}\!\big(\phi_l^{\text{rx}},\theta_l^{\text{rx}}\big)\,
    \qT_l(f)\,
    \qf_{\text{tx}}\!\big(\phi_l^{\text{tx}},\theta_l^{\text{tx}}\big) \\
    \cdot \exp\!\Big(j k \, \qr_{\text{rx},l}^{T}\qd_{\text{rx},u}\Big)\,
           \exp\!\Big(j k \, \qr_{\text{tx},l}^{T}\qd_{\text{tx},s}\Big),
\end{multline}
where $\qf_{\text{rx}}(\cdot)\!\in\!\bbC^{2\times 1}$ and $\qf_{\text{tx}}(\cdot)\!\in\!\bbC^{2\times 1}$ denote the antenna field patterns at the specified azimuth and elevation angles. The per-path transfer matrix $\qT_l(f)\!\in\!\bbC^{2\times 2}$ encapsulates all propagation effects along the $l$-th path. Here, $k = 2\pi/\lambda$ is the wavenumber, with $\lambda$ denoting the wavelength; $(\qr_{\text{rx},l},\qr_{\text{tx},l})$ are the unit direction vectors of arrival and departure, respectively, and $(\qd_{\text{rx},u},\qd_{\text{tx},s})$ represent the relative positions of the receive and transmit antenna elements with respect to their array centers.

In \eqref{eq:h_sul}, the angles of arrival/departure and their direction vectors $(\phi_l^{\text{rx}},\theta_l^{\text{rx}},\qr_{\text{rx},l})$ and $(\phi_l^{\text{tx}},\theta_l^{\text{tx}},\qr_{\text{tx},l})$ are determined by the transmitter-receiver geometry and surrounding environment. These parameters are frequency invariant, meaning that the multipath geometry observed at 3.5~GHz remains valid at 7~GHz.\footnote{The FR3 spectrum is expected to be deployed on the same grid as existing FR1 macro sites, enabling co-location and reuse of 5G infrastructure \cite{5GAmericas2024PurposeBackground}. Hence, even if base station locations differ slightly across frequency bands, they remain geographically close, and the small deterministic offsets can be accounted for by adjusting AoAs, AoDs, and delays.} 
The antenna field patterns $\qf_{\text{rx}}(\cdot)$ and $\qf_{\text{tx}}(\cdot)$ and the array geometry $(\qd_{\text{rx},u},\qd_{\text{tx},s})$ are frequency-specific but known a priori and can be incorporated into the model. The remaining frequency dependence is captured by the per-path transfer matrix $\qT_l(f)$, which accounts for path loss, material interaction, and polarization-dependent variations across frequency.

Therefore, low-band channels provide a reliable estimate of the underlying geometry (delays and angular parameters), whereas high-band channels can be inferred by reweighting these same paths with frequency-dependent complex coefficients. This decomposition underpins the learning framework: the model only needs to capture how $\qT_l(f)$ varies across frequencies while reusing the invariant multipath geometry.

\subsection{Ray-Tracing Dataset Generation} \label{RT}
To validate the above physical insight and provide data for learning, we construct paired 3.5~GHz and 7~GHz channel datasets using a systematic ray-tracing pipeline inspired by \cite{lin2025geo2commap}. The process begins with generating a building map and then performing ray-tracing simulations.

For a given area of dimensions $L_x \times L_y$, a building map $\qB \in \mathbb{R}^{N_x \times N_y}$ is generated with resolution $r$, where ${L_x}/{N_x} = {L_y}/{N_y} = r$. A 3D mesh model is then created using OpenStreetMap (OSM) \cite{openstreetmap} and Blender \cite{blender}: OSM provides building geometry data, while Blender converts it into detailed 3D models suitable for RT. 
Multipath propagation is simulated using Sionna \cite{aoudia2025sionna} and the Shooting and Bouncing Rays (SBR) algorithm. The simulations consider directional antennas at 3.5~GHz and 7~GHz, with boresight gains of 6.3~dBi and 12.3~dBi and half-power beamwidths of $65^\circ/8^\circ$ and $32.5^\circ/4^\circ$, respectively. Antenna orientations are randomized within $[0,2\pi)$ to capture diverse azimuthal conditions.

The SBR algorithm launches multiple rays from the transmitter, each of which undergoes specular reflection or scattering upon interacting with objects. Let $d_l$ denote the total length of path $l$ and $\tau_l = d_l/c$ its propagation delay. By assuming the transmitter and receiver are well-synchronized, the per-path transfer matrix can be modeled as \cite{aoudia2025sionna}
\begin{equation} \label{eq_alpha_l}
\qT_{l}(f) = \frac{\lambda}{4\pi d_l}\, e^{-j k d_l}\; \! \left(
\prod_{m=1}^{M_{l}} \mathbf{B}_{m}\,\mathbf{C}_{m}(f) \right) \!,
\end{equation}
where $\mathbf{C}_m(f)\in\!\bbC^{2\times 2}$ is the interaction coefficient (reflection or scattering), and $\mathbf{B}_m\in\!\bbR^{2\times 2}$ is the coordinate transformation matrix. Reflection and scattering behaviors follow Fresnel coefficients \cite{ITU2015} and stochastic scattering models \cite{aoudia2025sionna}.

In \eqref{eq_alpha_l}, the term $\lambda/(4\pi d_l)$ represents the free-space path loss, while material absorption and scattering efficiency further influence both the magnitude and phase through the interaction matrices $\mathbf{C}_m(f)$. The phase factor $e^{-j k d_l}$ accounts for the propagation delay, exhibiting a faster phase rotation at higher frequencies for a fixed $d_l$. Additional phase shifts are introduced by the reflection and scattering matrices $\mathbf{C}_m(f)$.

This procedure yields paired multipath features, including delays, angles, and per-path transfer matrices, at both frequencies, forming a realistic dataset that preserves the geometry-invariant structure while encoding frequency-dependent reweighting. These data serve as the foundation for the learning framework described next.

\subsection{Deep Learning for Cross-Band Mapping}
Building on the constructed dataset, we develop \textbf{CIR-UNext}, a deep learning framework designed to learn the mapping from 3.5~GHz to 7~GHz multipath channels. The framework comprises two specialized AU-Net models:  
\begin{itemize}
    \item \textbf{Gain-UNext}: an AU-Net-Aux shown in Fig.~\ref{fig:Gain-UNext} that takes 3.5~GHz path features as input and predicts the corresponding path gains at 7~GHz, represented as $|h_{s,u,l}(f)|$;  
    \item \textbf{Phase-UNext}: a standard AU-Net that takes path delays and interaction numbers as input and predicts the corresponding path phases at 7~GHz, denoted by $\angle h_{s,u,l}(f)$.  
\end{itemize}

Both models share a common backbone consisting of nine convolutional blocks and four downsampling/upsampling stages. Skip connections are used to preserve spatial features, while attention gates \cite{oktay2018attention} enhance focus on salient regions and improve convergence stability.  

\begin{figure}[t]
    \centering
    \subfloat[\label{Building Map 1}]{\includegraphics[width=0.35\linewidth]{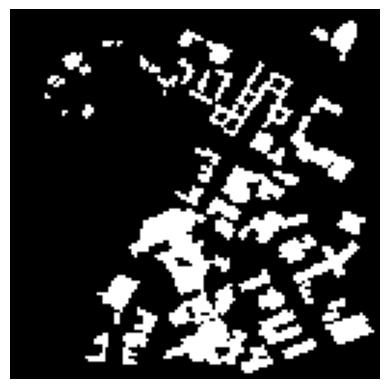}} 
    \hspace{0.05\linewidth}
    \subfloat[\label{Building Map 2}]{\includegraphics[width=0.35\linewidth]{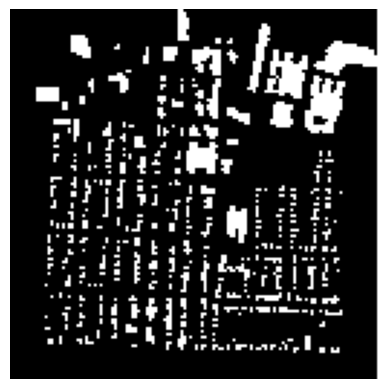}} 
    \caption{Building Maps used in the experiments. (a) BM1, characterized by a prominent open space. (b) BM2, featuring densely clustered buildings. Notably, both BM1 and BM2 are unseen during training.} 
    \label{fig:testing_buildings}
\end{figure}

The dataset is split at the building-map level into training, validation, and testing sets in an 80\%--10\%--10\% ratio. Rotation and mirroring are used for data augmentation to enhance robustness and generalization. The models are trained using mean square error (MSE) loss for 200 epochs on an NVIDIA RTX~5080 GPU. Testing is performed on 900 test samples, equally divided between two building scenarios: BM1 (open area) and BM2 (dense urban), as illustrated in Fig.~\ref{fig:testing_buildings}. 

With the overall framework established, the following sections analyze the two core components of \textbf{CIR-UNext}: Section~\ref{Gain-UNext} presents Gain-UNext, focusing on gain prediction and architectural optimization, while Section~\ref{Phase-UNext} introduces Phase-UNext, which models frequency-dependent phase variations.

\begin{table}[t]
\centering
\caption{Comparison of Average Absolute Error Statistics for Different Gain-UNext Architectures}
\label{tab:median_lqr}
\begin{tabular}{lcccc}
\hline
\textbf{Architecture} & \textbf{Q1} & \textbf{Q3} & \textbf{Median} & \textbf{IQR} \\
\hline
U-Net           & 0.43 (dB) & 2.03 (dB) & 1.01 (dB) & 1.60 (dB) \\  
AU-Net          & 0.32 (dB) & 1.82 (dB) & 0.81 (dB) & 1.50 (dB) \\
AU-Net-Parallel & 0.25 (dB) & 1.72 (dB) & 0.70 (dB) & 1.47 (dB)\\
\textbf{AU-Net-Aux}      & \textbf{0.28} (dB) & \textbf{1.61 (dB)} & \textbf{0.70 (dB)} & \textbf{1.34 (dB)} \\
\hline
\end{tabular}
\end{table}

\section{Gain-UNext} \label{Gain-UNext}
This section focuses on \textbf{Gain-UNext}, which is dedicated to predicting cross-band path gains. Building upon the physical motivation in Section~\ref{RT} and the dataset introduced earlier, we analyze how network architecture and input feature design influence the accuracy and robustness of gain prediction.  

\subsection{Different Gain-UNext Architectures}
We investigate four Gain-UNext architectures. The first is a standard U-Net, which serves as the baseline. The second is a standard AU-Net, which integrates attention gates (AGs) into the skip connections \cite{oktay2018attention}, enabling the network to emphasize salient regions and suppress irrelevant features.

The third architecture considered is a \emph{parallel-connected} AU-Net (AU-Net-Parallel), which predicts the multipath gains of line-of-sight (LoS) and non-line-of-sight (NLoS) components separately. Since LoS paths typically exhibit higher stability and stronger power, whereas NLoS paths are more sensitive to scattering and reflections, separating them enables the network to better capture the distinct statistical properties of each class of paths. 

The last architecture considered is AU-Net-Aux, which incorporates auxiliary information into the AU-Net framework to enhance prediction accuracy. In the input configuration, the interaction number (up to 5 interactions), horizontal AoD, and horizontal antenna pattern are all perfectly aligned with the target 7~GHz PG Map. We refer to these features as auxiliary information. In the baseline design, the encoder extracts features from all auxiliary information directly, which may lead to information confusion or loss during the feature extraction process. To address this issue, in AU-Net-Aux we not only feed the auxiliary information into the encoder to support feature extraction, but also downsample it and propagate it through the skip connections to the decoder, thereby assisting in map reconstruction. The overall AU-Net-Aux architecture is illustrated in Fig.~\ref{fig:Gain-UNext}.

To evaluate performance, we employ the absolute error over the entire area as the evaluation metric, defined as
\begin{equation}
\text{G-Err} =  \left| \hat{x}_{p,l} - x_{p,l} \right|,
\label{eq:diff_Gain-UNext_err}
\end{equation}
where $p$ denotes the valid positions (excluding building-occupied areas), $\hat{x}_{p,l}$ is the predicted path gain at the $p$-th position of the $l$-th path (corresponding to $|h_{s,u,l}(f)|$ in (\ref{eq:h_sul})), and $x_{p,l}$ is the corresponding ground-truth value. All path gains are expressed in decibels (dB).

Table~\ref{tab:median_lqr} summarizes the error statistics under a common input configuration illustrated in Fig.~\ref{fig:Gain-UNext} (excluding delay information). Standard U-Net produces the largest errors, with a median absolute error of $1.01$~dB and an interquartile range (IQR) of $1.60$~dB. Incorporating AGs, AU-Net reduces the median error to $0.81$~dB and the IQR to $1.50$~dB. AU-Net-Parallel further improves robustness by modeling LoS and NLoS paths separately, achieving a median error of $0.70$~dB (IQR $1.47$~dB), though at the cost of requiring two networks. AU-Net-Aux achieves the best overall performance, with the same median error of $0.70$~dB but a smaller IQR of $1.34$~dB, thus offering both higher accuracy and greater efficiency.

\begin{table}[t]
\centering
\caption{Comparison of Average Absolute Error Statistics for Different Input Configurations.}
\label{tab:input_lnfo}
\begin{tabular}{lcccc}
\hline
\textbf{Input Configuration} & \textbf{Q1} & \textbf{Q3} & \textbf{Median} & \textbf{IQR}\\
\hline
Baseline                     & 0.39 (dB) & 2.38 (dB) & 1.04 (dB) & 1.99 (dB) \\
+ Ant. pat.            & 0.37 (dB) & 1.79 (dB) & 0.95 (dB) & 1.79 (dB) \\
+ Ant. pat. + AoD      & 0.32 (dB) & 1.82 (dB) & 0.81 (dB) & 1.50 (dB) \\
\hline
\end{tabular}
\end{table}

\subsection{Useful Information for Cross-Band Path Gain Prediction}
This subsection investigates the impact of different input configurations for \textbf{Gain-UNext}. In addition to the frequency offset between input and output path gains, antenna patterns are also considered. To facilitate the learning of antenna pattern variations, we incorporate the horizontal antenna pattern defined in 3GPP TR~38.901 \cite{3gpp38901}:
\begin{equation} \label{eq_ant}
G(\phi) = G_{0} - \min \!{\left\{ 12 \left( \frac{\phi}{\phi_{3\text{dB}}} \right)^{2}, \; 30 \right\}},
\end{equation}
where $G_{0}$ is the boresight gain, $\phi_{3\text{dB}}$ is the half-power beamwidth, and $\phi$ is the angular offset from the antenna boresight. Since path gain is also strongly correlated with the AoD, including AoD as an additional input feature can further improve prediction performance.  

Table~\ref{tab:input_lnfo} summarizes the results under different input settings with the same standard AU-Net architecture. The Baseline configuration, using only 3.5~GHz path gain and interaction number, yields a median error of $1.04$~dB (IQR $1.99$~dB). Adding antenna pattern reduces the error to $0.95$~dB (IQR $1.79$~dB). Incorporating both antenna pattern and AoD further improves performance, lowering the median error to $0.70$~dB and the IQR to $1.34$~dB. These results confirm the effectiveness of antenna pattern and directional features when considering different antenna patterns in cross-band path gain prediction.

\subsection{Prediction Across Environments and Path Conditions}
\begin{figure}[t]
    \centering
    \subfloat[\label{fig:boxplot_diffBM}]{%
        \includegraphics[width=0.4\linewidth]{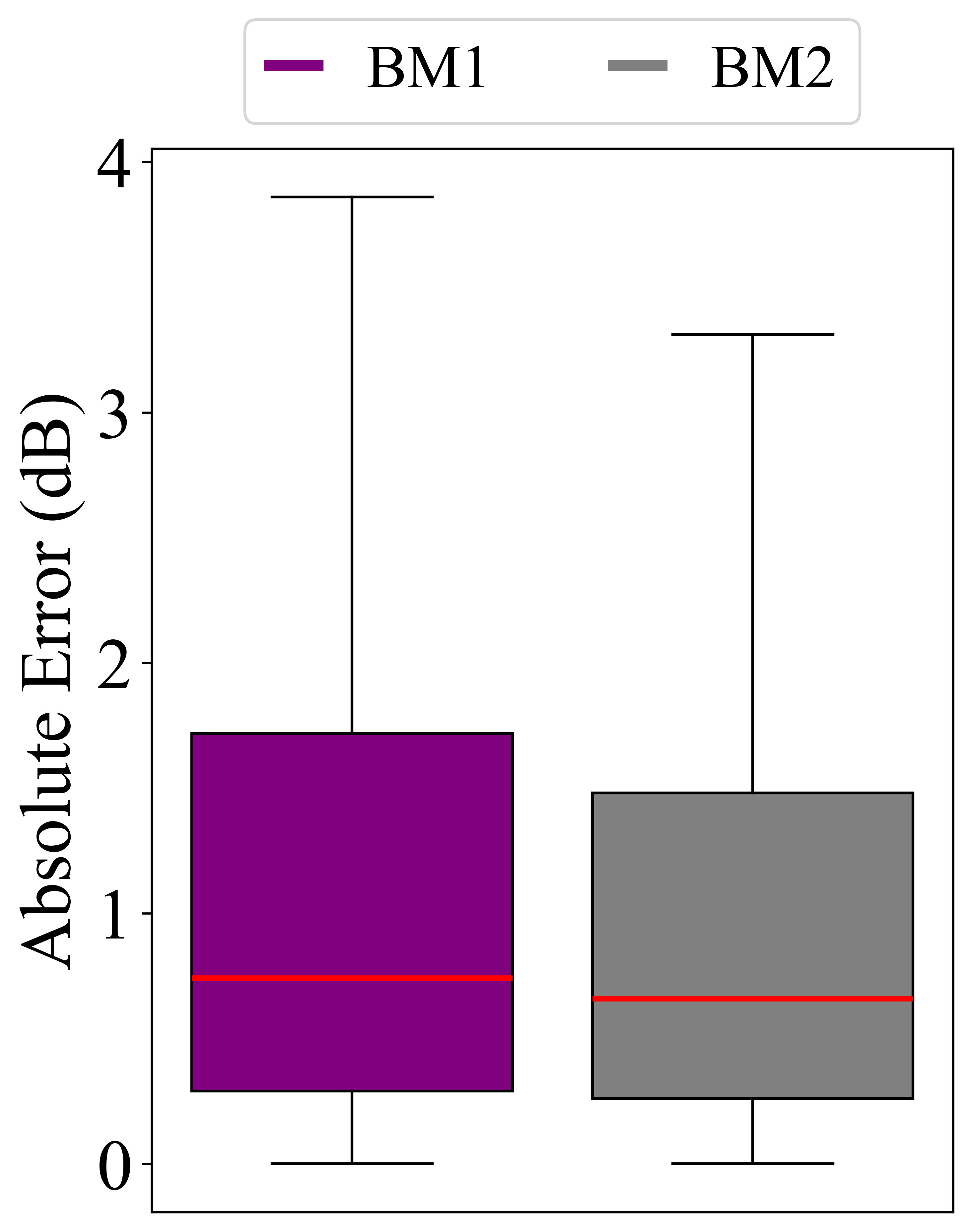}} 
    \hspace{0.05\linewidth}
    \subfloat[\label{fig:boxplot_delayinput}]{%
        \includegraphics[width=0.4\linewidth]{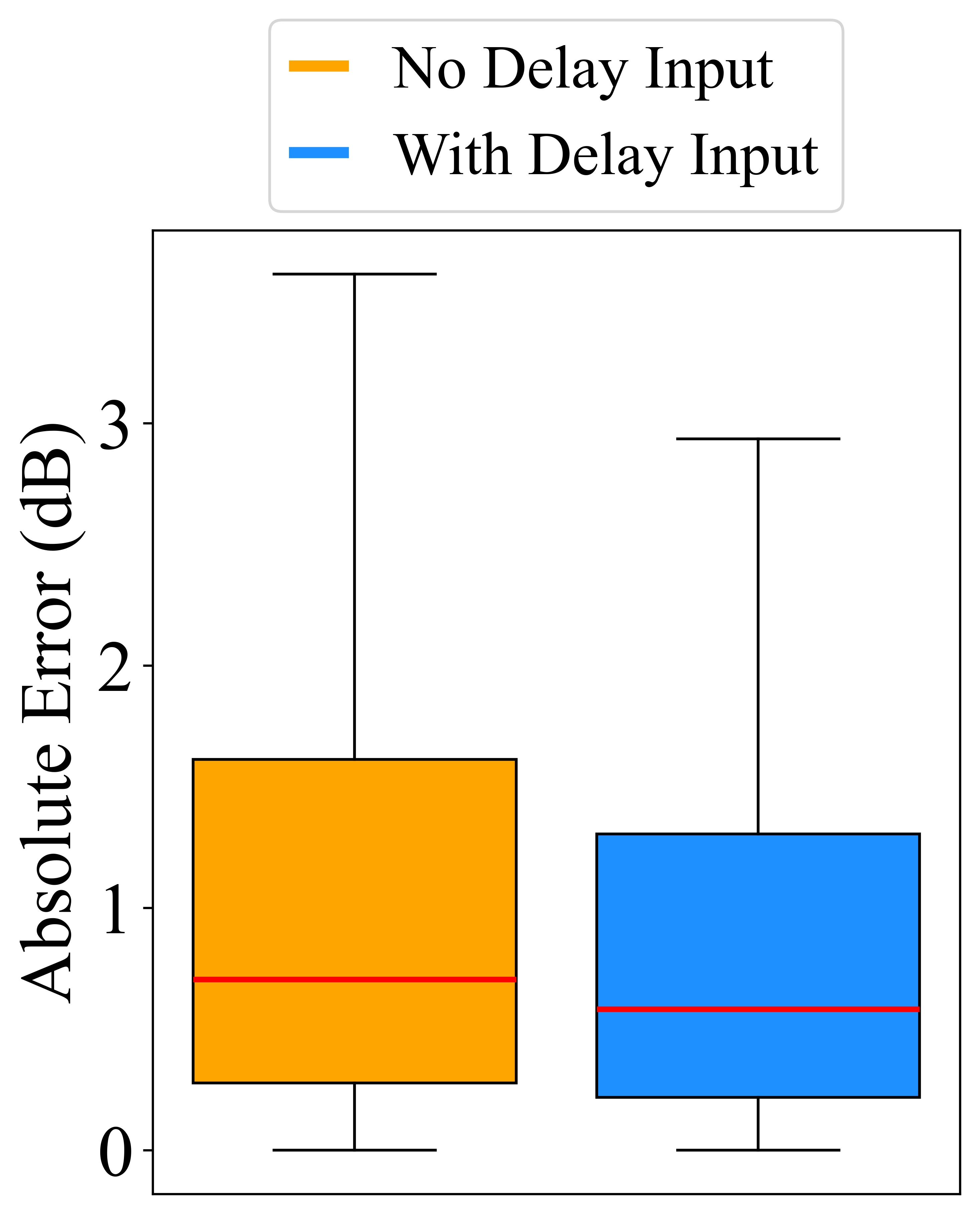}} 
    \caption{Boxplot comparison of path gain prediction errors: (a) between different building maps, and (b) with and without delay input (in both building maps). Each box represents the IQR of the absolute error, with the {\rl red} line inside indicating the median value. The whiskers extend to the furthest data points within the range defined as Q1$-1.5\cdot$IQR to Q3$+1.5\cdot$IQR, where Q1 and Q3 denote the first and third quartiles, respectively.}
    \label{fig:diffBM_delayinput}
\end{figure}

We evaluate the performance of \textbf{Gain-UNext} across different building maps (BMs), as shown in Fig.~\ref{fig:testing_buildings}. The AU-Net-Aux architecture is employed with 3.5~GHz path gain, interaction number, antenna pattern, and AoD as input features. Fig.~\ref{fig:boxplot_diffBM} compares the error distributions under different BMs. Gain-UNext performs better in BM2 than in BM1: in BM1, the median absolute error is $0.74$~dB (IQR $1.43$~dB), while in BM2, the median is $0.66$~dB (IQR $1.22$~dB).

\begin{figure}[t]
    \centering
    \subfloat[\label{High_Error}]{\includegraphics[width=0.45\linewidth]{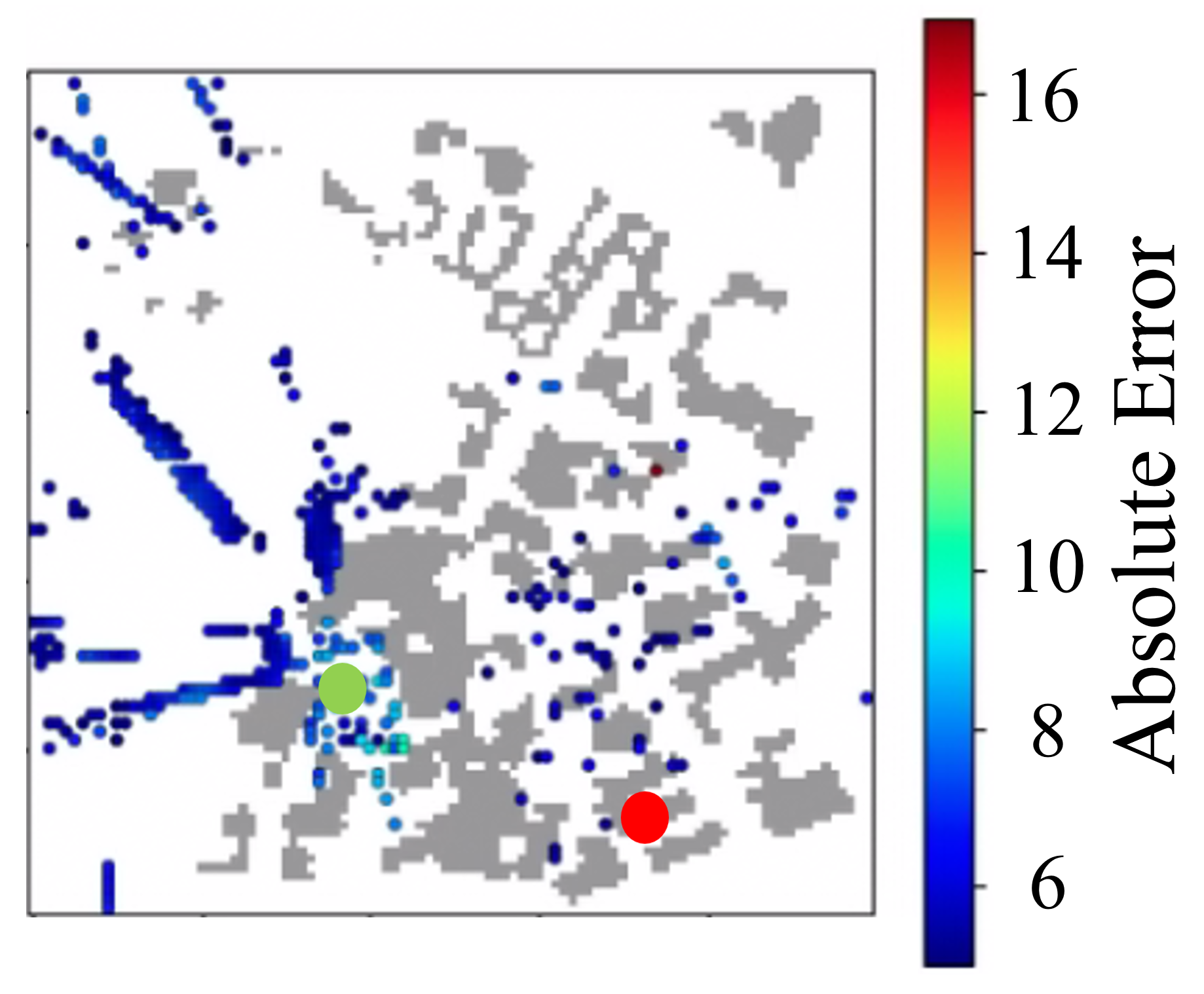}} 
    \hspace{0.05\linewidth}
    \subfloat[\label{High_Error_RT}]{\includegraphics[width=0.45\linewidth]{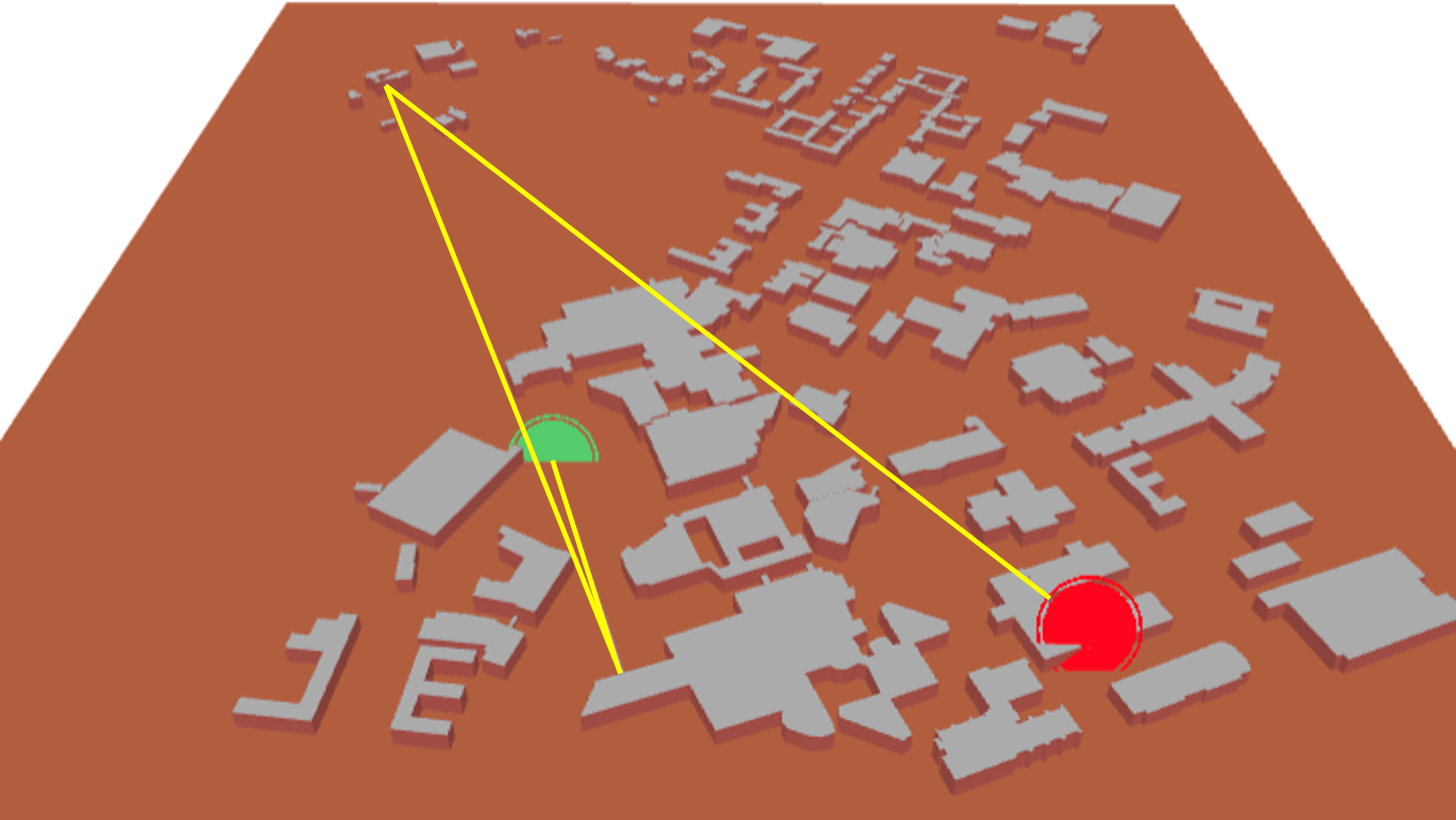}} 
    \caption{Regions of high error in BM1 and corresponding ray-tracing visualization. (a) Top view of top $5\%$ error distribution. The {\textcolor{red}{red}} dot denotes the BS, and the {\textcolor{green}{green}} dot denotes the UE. (b) Ray-tracing result of the BS-UE link in (a), where the {\textcolor{yellow}{yellow}} lines represent traced rays.} 
    \label{fig:High_Error_figs}
\end{figure}

To investigate the cause, we examine regions in BM1 with the largest errors and place user equipment (UE) within these areas. For each UE, the absolute error is calculated as $\left| \hat{x}_{i,j} - x_{i,j} \right|$, where $\hat{x}_{i,j}$ denotes the predicted first-path power and $x_{i,j}$ is the ground truth. Fig.~\ref{fig:High_Error_figs} shows the distribution of errors: Fig.~\ref{High_Error} highlights the top $5\%$ error regions in a top view, while Fig.~\ref{High_Error_RT} provides the corresponding ray-tracing visualization.  

\begin{table}[t]
\centering
\caption{Prediction Error Statistics under Different Interaction Numbers.}
\label{tab:interaction_error}
\begin{tabular}{ccccc}
\hline
\textbf{Interaction Number} & \textbf{Q1} & \textbf{Q3} & \textbf{Median} & \textbf{IQR} \\
\hline
LoS    & 0.09 dB & 0.45 dB & 0.21 dB & 0.36 dB \\
$I=1$  & 0.18 dB & 1.00 dB & 0.46 dB & 0.82 dB \\
$I=2$  & 0.31 dB & 1.40 dB & 0.71 dB & 1.09 dB \\
$I=3$  & 0.39 dB & 1.69 dB & 0.87 dB & 1.30 dB \\
$I=4$  & 0.46 dB & 1.93 dB & 1.01 dB & 1.47 dB \\
$I=5$  & 0.56 dB & 2.25 dB & 1.20 dB & 1.69 dB \\
\hline
\end{tabular}
\end{table}

The results indicate that the largest errors occur in regions without a LoS path, where the signal is dominated by reflections. As shown in Fig.~\ref{High_Error_RT}, although the geometric distance between the BS and UE is short, the actual propagation path length is much longer, leading to inaccurate predictions. Such paths occur more frequently in BM1, explaining its inferior performance compared to BM2. This is because BM1, characterized by larger open areas and fewer surrounding obstacles, often forces NLoS users to rely on long detour reflections from distant buildings. In contrast, BM2 contains denser building clusters that provide shorter and more stable reflection paths, which reduces the relative path length mismatch and improves prediction accuracy.

To mitigate this issue, delay information is incorporated into the input. As shown in Fig.~\ref{fig:boxplot_delayinput}, including delay significantly improves performance by enabling the model to better capture the true propagation distance and reducing errors in NLoS regions. Without delay, the median absolute error over the full test set is $0.70$~dB with an IQR of $1.34$~dB, whereas with delay the median decreases to $0.58$~dB and the IQR to $1.09$~dB.

We also analyze the prediction error of \textbf{Gain-UNext} under different numbers of interactions with the environment. The same architecture shown in Fig.~\ref{fig:Gain-UNext} is used. Table~\ref{tab:interaction_error} reports the error statistics. As expected, both the median and the spread of errors increase with the interaction number. This trend reflects the greater complexity of multipath propagation introduced by higher-order reflections, which makes accurate prediction increasingly challenging.

\section{Phase-UNext} \label{Phase-UNext} 
This section focuses on \textbf{Phase-UNext}, which is designed for cross-band path phase prediction. We first introduce the incorporation of propagation distance, induced phase shifts as auxiliary inputs to simplify the learning task. We then analyze the inherent challenges of predicting multipath phase using U-Net-like architectures.

\subsection{Propagation Distance Phase Shift as Input}
As described in Section~\ref{RT}, the complex transfer
matrix can be expressed as (\ref{eq_alpha_l}), where $e^{-j k d_l}$ represents the phase shift associated with the propagation distance $d_l$, and $\prod_{m=1}^{M_{l}} \mathbf{B}_{m}\mathbf{C}_{m}(f)$ accounts for the cumulative effects of multiple interactions on both path gain and phase.  

To simplify learning, instead of directly predicting the full frequency-dependent phase using the 3.5~GHz phase as input, we precompute the distance-dependent term $e^{-j k d_l}$ for 7 GHz and provide it as an input. This approach allows the network to focus solely on the residual phase variations caused by multipath interactions, thereby reducing model complexity and improving convergence stability.  

To evaluate performance, we employ the absolute phase error over an area as the evaluation metric, defined as
\begin{equation} 
\text{P-Err} = \min\!\left\{ |\hat\phi_{p,l}-\phi_{p,l}|,\; 2\pi-|\hat\phi_{p,l}-\phi_{p,l}| \right\},  \label{eq:P-Err}
\end{equation}
where $\hat{\phi}_{p,l}$ is the predicted value at the $p$-th position of the $l$-th path, and ${\phi}_{p,l}$ is the corresponding ground-truth value. 
The second term in \eqref{eq:P-Err} accounts for phase wrapping effects.

\begin{table}[t]
\centering
\caption{Comparison of average absolute error statistics between 3.5~GHz phase input and delay-based phase input across different interaction orders.}
\label{tab:phase_compare}
\begin{tabular}{c|cc|cc}
\hline
{\textbf{Case}} & \multicolumn{2}{c|}{\textbf{3.5~GHz Phase Input}} & \multicolumn{2}{c}{\textbf{Delay-Based Phase Input}} \\
 & IQR & Median & IQR & Median \\
\hline
LoS     & 0.11 (rad) & 0.06 (rad) & \textbf{0.08} (rad) & \textbf{0.05} (rad) \\
$I=1$   & 0.85 (rad) & 0.21 (rad) & \textbf{0.62} (rad) & \textbf{0.14} (rad) \\
$I=2$   & 1.57 (rad) & 0.65 (rad) & \textbf{0.95} (rad) & \textbf{0.27} (rad) \\
$I=3$   & 1.65 (rad) & 0.75 (rad) & \textbf{0.95} (rad) & \textbf{0.30} (rad) \\
$I=4$   & 1.66 (rad) & 0.78 (rad) & \textbf{0.85} (rad) & \textbf{0.29} (rad) \\
$I=5$   & 1.67 (rad) & 0.82 (rad) & \textbf{0.74} (rad) & \textbf{0.27} (rad) \\
\hline
\end{tabular}
\end{table}

Table~\ref{tab:phase_compare} compares the absolute error statistics obtained with the raw 3.5~GHz phase input and the proposed delay-based phase input. Results are reported separately for different interaction numbers ($I=1$ to $I=5$) and LoS cases.  
As shown in Table~\ref{tab:phase_compare}, the delay-based input consistently outperforms the direct 3.5~GHz phase input across all cases. For instance, when $I=5$, the median error decreases from $0.82$ to $0.27$~rad, and the IQR drops from $1.67$ to $0.74$~rad. These results demonstrate that precomputing the distance-dependent phase term significantly reduces the learning burden and improves phase prediction accuracy.

\subsection{Challenges of Phase Prediction}
While the phase of LoS paths can be accurately predicted, paths involving reflections or scattering remain considerably more difficult to model. As discussed in Section~\ref{RT}, both reflection and scattering contribute to complex, nonlinear phase variations along the propagation path.

For reflections, the key influencing factors are the complex relative permittivity $\eta$ and the incidence angle $\theta_0$.
While $\eta$ mainly depends on carrier frequency and material properties and is thus stable, the absence of explicit $\theta_0$ information in the current Phase-UNext architecture makes phase prediction difficult.
Providing all incidence angles as inputs would drastically increase dimensionality, approaching the complexity of directly solving the full ray-tracing equations, defeating the purpose of a neural approximation.
For scattering, random phase shifts $\chi_1$ and $\chi_2$ are inherently stochastic and cannot be effectively learned by a deterministic, data-driven network..

Taken together, these factors explain why U-Net-like architectures face intrinsic limitations in modeling multipath phase. Without explicit geometric details (e.g., incidence angles) and with stochastic scattering effects, the network cannot fully capture the underlying physical dependencies, which constrains the achievable prediction accuracy.

\section{Channel2ComMap}
As an application, this section presents \textbf{Channel2ComMap}, which extends \textbf{CIR-UNext} as a foundation model for throughput (Tput) prediction in MIMO-OFDM systems. Building upon the predicted multipath channel features from Gain-UNext, \textbf{Channel2ComMap} demonstrates how cross-band channel prediction can directly benefit downstream communication performance estimation.

The proposed deep learning model adopts a standard AU-Net architecture, similar as shown in Fig. \ref{fig:Gain-UNext}, with a $(6\times20)$-channel input formulated as  
\begin{equation}
    \Big[\hat{\qP}, \underbrace{\boldsymbol{\tau}, \mathbf{AoD}_{\text{h}}, \mathbf{AoD}_{\text{v}}, \mathbf{AoA}_{\text{h}}, \mathbf{AoA}_{\text{v}}}_{\text{auxiliary~features}}\Big] 
    \in \mathbb{R}^{N_x \times N_y \times 120},
\end{equation}
where $\hat{\qP} \in \mathbb{R}^{N_x \times N_y \times 20}$ denotes the predicted multipath gain map from Gain-UNext,  
$\boldsymbol{\tau} \in \mathbb{R}^{N_x \times N_y \times 20}$ represents the path delay map,  
$\mathbf{AoD}_{\text{h}}$ and $\mathbf{AoD}_{\text{v}}$ correspond to the horizontal and vertical AoD,  
and $\mathbf{AoA}_{\text{h}}$ and $\mathbf{AoA}_{\text{v}}$ correspond to the horizontal and vertical angles of arrival (AoA). 
Based on these inputs, the model predicts a Tput map that captures the spatial distribution of achievable rates across the target area.  
We also evaluate an enhanced variant, AU-Net-Aux, as shown in Fig. \ref{fig:Gain-UNext} in which the auxiliary features are propagated through skip connections to improve reconstruction fidelity.

\begin{figure}[t]
    \centering
    \includegraphics[width=0.75\linewidth]{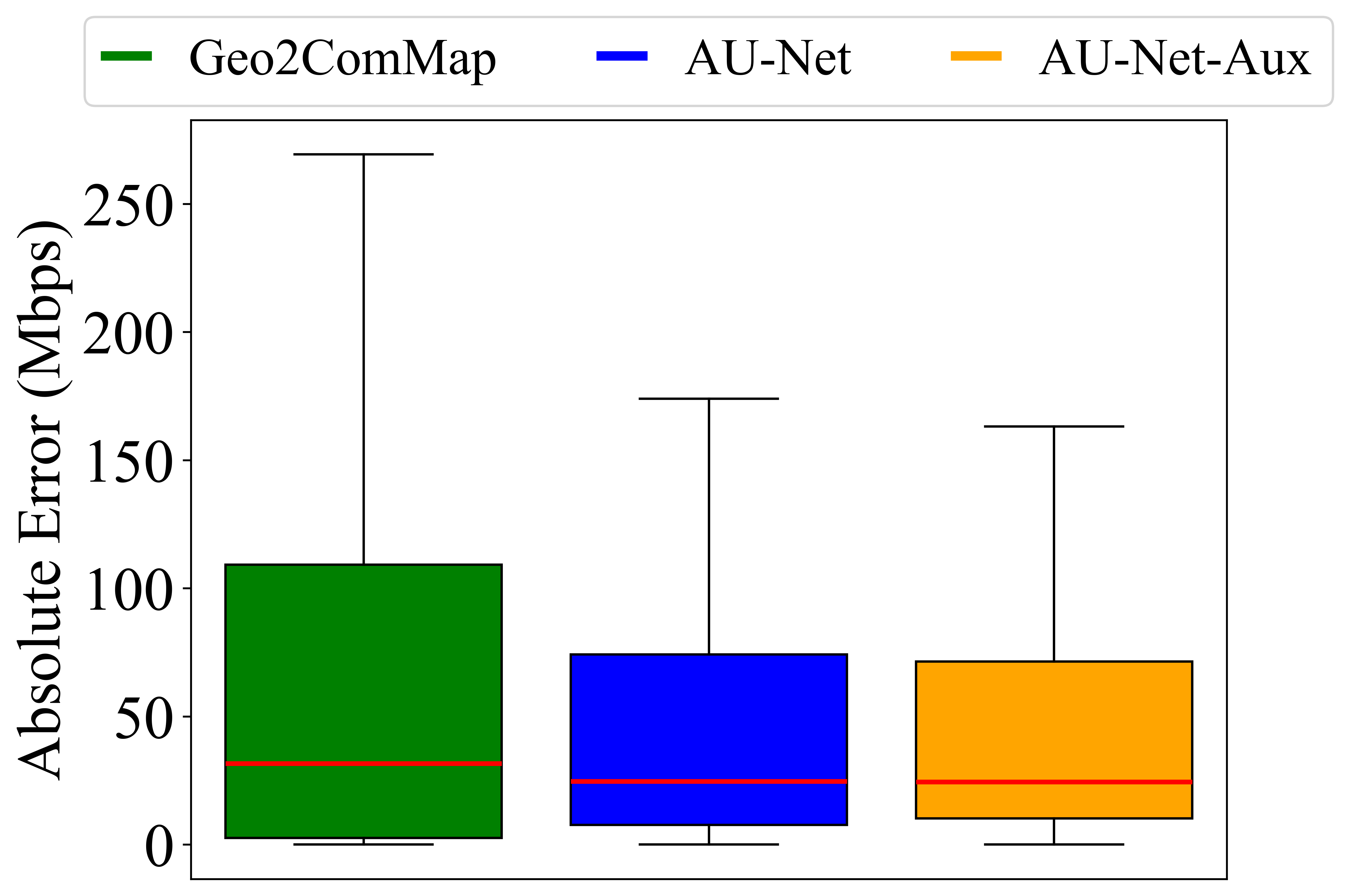}
    \caption{Boxplot comparison of Tput prediction errors for Channel2ComMap and Geo2ComMap. AU-Net represents Channel2ComMap with the standard AU-Net backbone, while AU-Net-Aux denotes Channel2ComMap with the AU-Net-Aux architecture.}
    \label{fig:boxplot_tput}
\end{figure}

Prior work, such as \textbf{Geo2ComMap} \cite{lin2025geo2commap}, considers Tput, rank indicator (RI), and channel quality indicator (CQI) as input features. In contrast, \textbf{Channel2ComMap} utilizes the multipath channel features predicted by \textbf{CIR-UNext} to produce fine-grained Tput maps. Using the same MIMO-OFDM configuration as \cite{lin2025geo2commap}, we compare the two frameworks, and Fig.~\ref{fig:boxplot_tput} presents their prediction error distributions over a 0--1900~Mbps range.
Geo2ComMap exhibits a median error of $31.64$~Mbps with an IQR of $106.68$~Mbps.  
Channel2ComMap with the standard AU-Net reduces the median error to $24.70$~Mbps and the IQR to $66.55$~Mbps.  
The AU-Net-Aux variant achieves the best overall performance, further reducing the median error to $24.49$~Mbps and the IQR to $61.16$~Mbps.  
These results demonstrate that incorporating predicted multipath channel features enables more accurate throughput prediction, effectively bridging the gap between physical-layer sensing and link-layer performance estimation.

\section{Conclusion} \label{Conclusion}
This work presented \textbf{CIR-UNext}, a deep learning framework for cross-band channel prediction that transfers multipath information from 3.5~GHz to 7~GHz. Using ray-tracing datasets and specialized AU-Net variants, the framework accurately predicts path gain across frequencies. The AU-Net-Aux model achieves a median gain error of 0.58~dB, while incorporating distance-dependent phase terms reduced the median phase error to 0.27~rad even for fifth-order interactions.  
Furthermore, \textbf{Channel2ComMap} extends CIR-UNext as a foundation model for MIMO-OFDM throughput prediction, achieving a median error of 24.70~Mbps over a 0--1900~Mbps range. These results highlight the potential of cross-band learning to bridge physical-layer channel modeling and link-layer performance estimation for future 6G networks.

\section*{Acknowledgment}
This work was supported in part by the National Science and Technology Council (NSTC) of Taiwan under Grants NSTC 114-2221-E-110-031-MY3, NSTC 114-2218-E-110-005, and NSTC 113-2222-E-002-010, and in part by the SNS JU project 6G-DISAC under the EU’s Horizon Europe research and innovation programme under Grant Agreement No. 101139130. 
\bibliographystyle{IEEEtran}
\bibliography{references-short}

\end{document}